\overfullrule=0pt
\input harvmac
\def\a{{\alpha}}

\def\l{{\lambda}}
\def\b{{\beta}}

\def\g{{\gamma}}
\def\k{{\kappa}}

\def\d{{\delta}}

\def\e{{\epsilon}}
\def\s{{\sigma}}

\def\half{{1\over 2}}
\def\p{{\partial}}

\def\t{{\theta}}

\Title{\vbox{\hbox{IFT-P.039/2001 }}}
{\vbox{
\centerline{\bf Covariant Quantization of the Superparticle using Pure
Spinors}}}
\bigskip\centerline{Nathan Berkovits\foot{e-mail: nberkovi@ift.unesp.br}}
\bigskip
\centerline{\it Instituto de F\'\i sica Te\'orica, Universidade Estadual
Paulista}
\centerline{\it Rua Pamplona 145, 01405-900, S\~ao Paulo, SP, Brasil}

\vskip .3in

The ten-dimensional
superparticle is covariantly quantized by constructing
a BRST operator from the fermionic Green-Schwarz constraints
and a bosonic pure spinor variable. This same method
was recently used for covariantly quantizing the superstring,
and it is hoped that the simpler case of the superparticle will
be useful for those who want to study this quantization method.
It is interesting that quantization of the superparticle action
closely resembles quantization of the worldline action for 
Chern-Simons theory.

\Date {May 2001}

\newsec{Introduction}

Recently, the ten-dimensional
superstring was covariantly quantized by constructing
a BRST operator from the fermionic Green-Schwarz constraints
and a bosonic pure spinor variable \ref\me
{N. Berkovits, JHEP 04 (2000) 018, hep-th/0001035\semi
N. Berkovits, hep-th/0008145.}. Although this
method was successfully used for computing tree-level scattering amplitudes,
the construction of the BRST operator is 
non-conventional so it is a bit mysterious why this method works.
This same BRST construction can be used
for covariantly quantizing the ten-dimensional
superparticle and it is hoped that by studying
this simpler model, some of the mysteries will be 
easier to understand.
 
Since the spectrum of the ten-dimensional superparticle 
contains a spin-one field, the constraints of the worldline action
should imply spacetime gauge invariances as well as spacetime
equations of motion. This differs from the worldline actions for
the particle or spinning particle where the constraints imply the Klein-Gordon
or Dirac equations of motion,
but do not imply spacetime gauge invariances.
One worldline action which does describe a theory with spacetime
gauge invariance is the worldline
version of Witten's action for Chern-Simons theory \ref\cs{E. Witten,
hep-th/9207094.}. It will turn out that the constraints and quantization
of this Chern-Simons 
action closely resemble the constraints and quantization
of the pure spinor version of the superparticle action \foot{The similarity
of the two worldline actions was first pointed out to me by Warren Siegel.}. 

Section 2 of this paper will review the problems with
quantizing the standard superparticle action. In section 3, the
worldline action for Chern-Simons will be discussed. Section 4
will review the superspace description of ten-dimensional super-Yang-Mills.
And in section 5, the pure spinor version of the superparticle action
will be quantized in a manner similar to the Chern-Simons action.
The appendix will contain a computation of the zero-momentum BRST cohomology
of the superparticle.

\newsec{Review of Standard Superparticle Description}

The standard action for the ten-dimensional superparticle is 
\ref\super{L. Brink and J.H. Schwarz, Phys. Lett. 100B (1981) 310.}
\eqn\action{
S=\int d\tau (\Pi^m P_m + e P^m P_m)}
where 
\eqn\defpi{\Pi^m = \dot x^m - {i\over 2} \dot\t^\a\g^m_{\a\b}\t^\b,}
$P_m$ is the canonical momentum for $x^m$, and $e$
is the Lagrange multiplier which enforces the mass-shell condition.
The gamma matrices $\g^m_{\a\b}$ and $\g_m^{\a\b}$ are
$16\times 16$ symmetric matrices which satisfy
$\g^{(m}_{\a\b} \g^{n)~ \b\g}= 2 \eta^{mn} \d_\a^\g$.
In the Weyl representation, 
$\g^m_{\a\b}$ and $\g_m^{\a\b}$ are
the off-diagonal blocks of the $32\times 32$ $\Gamma^m$ matrices.

The action of \action\ is spacetime-supersymmetric under
$$\d \t^\a = \e^\a, \quad x^m = {i\over 2}\t\g^m\e, \quad \d P_m = \d e =0,$$
and is also invariant under the local $\kappa$ transformations
\ref\sieg{W. Siegel, Phys. Lett. 128B (1983) 397.}
\eqn\ka{ \d \t^\a = P^m (\g_m \k)^\a,\quad
\d x^m = -{i\over 2}\t\g^m\d\t ,\quad \d P_m =0,\quad \d e= i\dot\t^\b\k_\b.}
The canonical momentum to $\t^\a$, which will be called $p_\a$, satisfies
$$ p_\a = \d L/ \d\dot\t^\a =-{i\over 2} P^m (\g_m\t)_\a,$$
so
canonical
quantization requires that physical states are annihilated by the
fermionic Dirac constraints defined by
\eqn\dirac{d_\a = p_\a +{i\over 2} P_m (\g^m\t)_\a.}
Since $\{p_\a,\t^\b\}=-i\d_\a^\b$,
these constraints satisfy the Poisson brackets 
\eqn\antic{\{d_\a, d_\b\} =  P_m \g^m_{\a\b},}
and since $P^m P_m =0$ is also a
constraint, eight of the sixteen Dirac constraints
are first-class and eight are second-class.
One can easily check that the eight first-class Dirac constraints
generate the $\kappa$ transformations of \ka, however, there
is no simple way to covariantly separate out the second-class
constraints. 

Nevertheless, one can easily quantize the superparticle in a non-Lorentz
covariant manner and obtain the physical spectrum.
Assuming non-zero $P^+$,
the local fermionic $\k$-transformations can be used to
gauge-fix $(\g^+\t)_\a =0$ where
$\g^\pm = {1\over{\sqrt{2}}}
(\g^0 \pm \g^9)$. In this gauge, the action of \action\
simplifies to the quadratic action \ref\lc{M.B. Green and
J.H. Schwarz, Nucl. Phys. B181 (1981) 502.}
\eqn\lcaction{S=\int d\tau (\dot x^m P_m +{i\over 2}
 P^+ (\dot\t \g^- \t) + e P^m P_m)
=
\int d\tau (\dot x^m P_m + {i\over 4}\dot\sigma_a \sigma_a + e P^m P_m) ,}
where $\sigma_a = \sqrt{2 P^+} (\g^-\t)_a$ and
$a=1$ to 8 is an $SO(8)$ chiral spinor index.

Canonical quantization of \lcaction\ implies that
$\{\sigma_a, \sigma_b \} = 2\d_{ab}$. So $\sigma_a$ 
acts like a `spinor' version of
the $SO(8)$ Pauli matrices
$\s^j_{a\dot b}$ which satisfy
$$\s^j_{a \dot c} 
\s^j_{b \dot d} + 
\s^j_{b \dot c} 
\s^j_{a \dot d} = 2 \d_{ab} \d_{\dot c \dot d}$$ 
where $j$ and $\dot b$ are $SO(8)$ vector and antichiral spinor indices.
One can therefore define the quantum-mechanical wavefunction $\Psi(x)$
to carry either an $SO(8)$ vector index, $\Psi_j(x)$,
or an $SO(8)$ antichiral spinor index, $\Psi_{\dot a}(x)$, and the
anticommutation relations of $\sigma_a$ are reproduced by defining
\eqn\wfc{\s^a \Psi_j(x) = \sigma_j^{a\dot b} \Psi_{\dot b}(x),\quad
\s_a \Psi_{\dot b}(x) = \sigma^j_{a\dot b} \Psi_j (x).}
Furthermore, the constraint $P_m P^m$ implies the linearized
equations of motion $\p_m\p^m \Psi_j =
\p_m\p^m \Psi_{\dot b} =0$.

So the physical states of the superparticle are described by a massless
$SO(8)$ vector $\Psi_j(x)$ and a massless
$SO(8)$ antichiral spinor $\Psi_{\dot a}(x)$ which are the physical
states of $d=10$ super-Yang-Mills theory. However, this description of
super-Yang-Mills theory only manifestly
preserves an $SO(8)$ subgroup of the super-Poincar\'e
group, and one would like a more covariant method for
quantizing the theory.
Covariant quantization can be extremely useful if one wants
to compute more than just
the physical spectrum in a flat background. For example, non-covariant  
methods are extremely clumsy for computing scattering amplitudes or
for generalizing to curved backgrounds.

Since the super-Yang-Mills spectrum contains a massless vector, one expects
the covariant superparticle constraints to generate the spacetime
gauge invariances of this vector. Note that these constraints
are not present in the gauge-fixed action of \lcaction\ since 
$\Psi_j$ describes only the transverse degrees of freedom
of the $SO(9,1)$ vector. Before describing the covariant
constraints which generate the gauge invariances of this vector,
it will be useful to first review the worldline action for Chern-Simons
theory which also has constraints related to spacetime gauge invariances.

\newsec{Worldline Description of Chern-Simons Theory}

Since the gauge invariance of a massless vector field
is $\d A_\mu = \p_\mu \Lambda$, one might guess that the
worldline action for such a 
field should contain the constraints $P_\mu.$
Although these constraints are too strong for describing Yang-Mills
theory, they are just right for describing $d=3$ Chern-Simons theory where
the field-strength of $A_\mu$ vanishes on-shell.

\subsec{Worldline action}

As was shown in \cs, Chern-Simons theory can be described using the
worldline action
\eqn\csaction{
S=\int d\tau (\dot x^\mu P_\mu + l^\mu P_\mu)}
where $\mu=0$ to 2 and $l^\mu$ are Lagrange multipliers for the constraints.
Since the constraints are first-class, the action can be quantized
using the BRST method. After gauging $l^\mu=-\half P^\mu$, the gauge-fixed
action is 
\eqn\csgf{
S=\int d\tau (\dot x^\mu P_\mu -\half P^\mu P_\mu
+ \dot c^\mu b_\mu)}
with the BRST operator
\eqn\csbrst{Q = c^\mu P_\mu}
where $(c^\mu,b_\mu)$ are fermionic Fadeev-Popov ghosts and anti-ghosts.

To show that the cohomology of the BRST operator describes Chern-Simons
theory, note that the most general wavefunction
constructed from a ground state
annihilated by $b^\mu$ is 
\eqn\csfield{\Psi(c,x) = C(x) + c^\mu A_{\mu}(x) + {i\over 2}\e_{\mu\nu\rho}
c^\mu c^\nu A^{*\rho}(x) + {i\over 6}
\e_{\mu\nu\rho}c^\mu c^\nu c^\rho C^*(x)}
where the expansion in $c^\mu$ terminates since $c^\mu$ is fermionic.
One can check that 
\eqn\qphi{Q\Psi= -i c^\mu \p_\mu C -{i\over 2} c^\mu c^\nu \p_{[\mu} A_{\nu]}
+ {1\over 6} 
\e_{\mu\nu\rho}c^\mu c^\nu c^\rho \p_\sigma A^{*\sigma}(x).}
So $Q\Psi=0$ 
implies that
$A_\mu(x)$ satisfies the equations of motion $\p_{[\mu} A_{\nu]}=0$
which is the linearized
equation
of motion of the Chern-Simons field. Furthermore,
if one defines the gauge parameter
$\Omega(c,x) = i\Lambda(x) - c^\mu \omega_{\mu}(x) + ...$, the gauge
transformation $\delta\Psi=Q\Omega$ implies $\d A_\mu = \p_\mu\Lambda$
which is the linearized gauge transformation of the Chern-Simons field.
Similarly,
$Q\Psi=0$ and $\delta\Psi= Q\Omega$ implies that $A^{*\rho}$ satisfies
the equation of motion $\p_\sigma A^{*\sigma}=0$ with the gauge invariance
$\d A^{*\sigma} = \e^{\sigma\mu\nu}\p_{\mu} w_{\nu}$, which are the
linearized equations of motion and gauge invariance of the Chern-Simons
antifield. The remaining fields, $C(x)$ and $C^*(x)$, describe
the spacetime ghost and antighost of Chern-Simons theory.

These equations of motion and gauge invariances can be obtained from
the Batalin-Vilkovisky version \ref\BV{I.A. Batalin and G.A.
Vilkovisky, Phys. Rev. D28 (1983) 2567.} of the
abelian Chern-Simons spacetime action
\eqn\css{{\cal S} = \int d^3 x (\half \e^{\mu\nu\rho} A_\mu \p_\nu A_\rho
+ i A^{*\mu} \p_\mu C),}
where, in addition to the usual Chern-Simons action for $A_\mu$,
there is a term coupling the antifield $A^{*\mu}$ to the gauge variation
of $A_\mu$. The action of \css\ can be written compactly in terms of the
wavefunction $\Psi$ of \csfield\ as 
\eqn\comp{{\cal S} = \half\int d^3 x \langle \Psi Q \Psi \rangle}
where $\langle ~~\rangle$ is normalized such that $\langle
c^\mu c^\nu c^\rho \rangle = i\e^{\mu\nu\rho}$.

\subsec{Non-abelian Chern-Simons theory}

Up to now, only abelian Chern-Simons theory has been discussed,
but it is easy to generalize to the non-abelian case. For example, the
Batalin-Vilkovisky version of the non-abelian Chern-Simons action is
\eqn\csna{{\cal S} = Tr \int d^3 x (\e^{\mu\nu\rho} (\half
A_\mu \p_\nu A_\rho + {i g\over 3} A_\mu A_\nu A_\rho) }
$$+
i A^{*\mu} (\p_\mu C + ig [A_\mu, C]) - g C C C^* ),$$
which can be written compactly as
\eqn\compna{{\cal S} = Tr\int d^3 x \langle \half\Psi Q \Psi
+{g\over 3} \Psi \Psi \Psi \rangle}
where $g$ is the Chern-Simons coupling constant and the fields in
$\Psi$ of \csfield\ now carry Lie algebra indices.\foot{It is interesting
to note that the Chern-Simons action can also be written in
manifestly gauge-invariant notation as 
${\cal S} = \half Tr\int_{\cal M}
d^4 x \langle (Q \Psi + g\Psi\Psi)^2 \rangle_4 $
where ${\cal M}$ is a four-dimensional volume with a three-dimensional
boundary at $x_3=0$,
$Q= c^M P_M$ for $M=0$ to 3, $\langle c^M c^N c^P c^Q \rangle_4 = \e^{MNPQ}$,
and $\Psi$ now depends on $x^3$ and $c^3$. Using cyclicity of the trace,
one finds ${\cal S}= Tr\int_{\cal M} d^4 x \langle Q(\half\Psi Q\Psi +
{g\over 3}\Psi\Psi\Psi)\rangle_4 = Tr\int d^4 x \langle \d(c^3)\d(x^3)
(\half\Psi Q\Psi +
{g\over 3}\Psi\Psi\Psi)\rangle_4$
since $\int_{\cal M} d^4 x
\langle Q \Lambda \rangle_4 $ only gets contributions from
the three-dimensional boundary at $x_3=0$. 
So ${\cal S}$ coincides with the Chern-Simons action defined in \compna.}
Note that the non-linear equations of motion and gauge invariances associated
with this action are
\eqn\nonlin{Q\Psi + g\Psi \Psi =0,\quad \d\Psi = Q\Omega +g [\Omega,\Psi].}

To construct a wordline action for non-abelian Chern-Simons theory
with $SO(N)$ gauge group, one introduces $N$ real fermionic variables
$\eta_I$ for $I=1$ to $N$ and modifies the worldline
action of \csgf\ to\foot{For $U(N)$ gauge group,
one introduces $N$ complex fermionic variables $(\eta_I,\overline\eta^I)$ 
with the action
$-i\int d\tau~\overline\eta^I\nabla\eta_I.$}
\eqn\nonac{
S=\int d\tau (\dot x^\mu P_\mu -\half P^\mu P_\mu
-{i\over 2} \eta_I \nabla \eta_I
+\dot c^\mu b_\mu)}
where 
$\nabla\eta_I = \dot\eta_I + g \eta_J \dot x^\mu
\tilde A^{IJ}_\mu(x) $ and $\tilde A^{IJ}_\mu(x)=
-\tilde A^{JI}_\mu(x)$ is a non-abelian
Chern-Simons background field \ref\fw{D. Friedan
and P. Windey, Nucl. Phys. B235 (1984) 395.}. 
Note that the constraints $P_\mu$
are conserved when the background is on-shell since
\eqn\onsh{\dot P_\mu = {{ig}\over 2}( {\p\over{\p\tau}}
(\eta_I \eta_J \tilde A^{IJ}_\mu) -  \eta_I \eta_J 
\dot x^\nu \p_\mu \tilde A^{IJ}_\nu )}
$$= -{{ig}\over 2}\eta_I \eta_J \dot x^\nu (\p_{[\mu} \tilde A_{\nu]}^{IJ}
+g \tilde A^{IK}_{[\mu} \tilde A^{KJ}_{\nu]} )=0.$$

After gauge-fixing, the BRST charge is still $Q=c^\mu P_\mu$,
but because of the background gauge field $\tilde A^{IJ}_\mu$
in \nonac, the canonical
momentum for $x^\mu$ is now ${{\p L}\over{\p \dot x^\mu}} =
P_\mu -{{ig}\over 2} \eta_I \eta_J \tilde A^{IJ}_\mu$. Since
$\{\eta_I,\eta_J\}=\d_{IJ}$,
\eqn\nonaq{Q (\eta_I \eta_J \Psi^{IJ}) = 
\eta_I \eta_J ( -i c^\mu (\nabla_\mu C)^{IJ} -{i\over 2}
c^\mu c^\nu (\nabla_{[\mu} A_{\nu]})^{IJ} +{1\over 6} \e_{\mu\nu\rho}
c^\mu c^\nu c^\rho (\nabla_\sigma A^{*\sigma})^{IJ})}
where 
$(\nabla_\mu s)^{IJ} = \p_\mu s^{IJ} + g (s^{IK}\tilde A_\mu^{KJ} -
s^{JK}\tilde A_\mu^{KI} )$ and 
$\Psi^{IJ}$ is defined as in \csfield.
So in a background gauge field $\tilde A_\mu^{IJ}$, the linearized
equations of motion of the non-abelian Chern-Simons
field and antifield are correctly described by
$Q (\eta_I \eta_J \Psi^{IJ}) = 0.$ 

Using intuition learned from this worldline description of 
Chern-Simons theory, it will be shown how to quantize the superparticle
in a similar manner. However, before performing this quantization, 
it will be useful to first review the superspace description of
ten-dimensional super-Yang-Mills theory.

\newsec{ Covariant Description of Super-Yang-Mills Theory}

Although on-shell super-Yang-Mills theory can be described by the 
$SO(8)$ wavefunctions $\Psi_j(x)$
and $\Psi_{\dot a}(x)$ of \wfc\ satisfying the linearized
equations of motion $ \p_m\p^m \Psi_j =
\p_m\p^m \Psi_{\dot a} =0,$ there are more covariant descriptions 
of the theory. Of course, there is a Poincar\'e-covariant description
using an $SO(9,1)$ vector field $a_m(x)$ and an $SO(9,1)$
spinor field $\chi^\a(x)$ transforming in the adjoint representation
of the gauge group which satisfy the equations of motion 
\eqn\lorentz{
\p^m f_{mn} + ig [a^m,f_{mn}]=0,\quad  \g^m_{\a\b}(\p_m \chi^\b + ig
[a_m,\chi^\b])=0,}
and gauge invariance 
\eqn \gaui{\d a_m = \p_m s + ig [a_m,s],\quad
\d \chi^\a =  ig [\chi^\a,s], \quad \d f_{mn} = ig [f_{mn},s],}
where $f_{mn}= \p_{[m} a_{n]} + ig [a_m,a_n]$ is the Yang-Mills
field strength and $g$ is the super-Yang-Mills
coupling constant. 
However, there is also a super-Poincar\'e covariant
description 
using an $SO(9,1)$ spinor wavefunction $A_\a (x,\t)$ defined
in $d=10$ superspace. As will be explained below, on-shell super-Yang-Mills
theory
can be described by a spinor superfield 
$A_\a (x,\t)$ transforming in the adjoint representation
which satisfies the superspace equation
of motion\ref\supers{W. Siegel, Phys. Lett. 80B (1979) 220.}
\eqn\eom{\g_{mnpqr}^{\a\b} (D_\a A_\b +i g A_\a A_\b) = 0}
for any five-form direction $mnpqr$,
with the gauge invariance
\eqn\superg{\d A_\a =  D_\a \Lambda + ig [A_\a,\Lambda]}
where $\Lambda(x,\t)$ is any scalar superfield and 
$$D_\a = {\p\over{\p\t^\a}} + {i\over 2}(\g^m\t)_\a \p_m$$
is the supersymmetric derivative.

One can also define field strengths constructed from $A_\a$ by
\eqn\fs{B_m = -{i\over 8}\g_m^{\a\b} (D_\a A_\b +i g A_\a A_\b),\quad
W^\a = {1\over{10}}\g_m^{\a\b} (D_\a B^m - \p^m A_\a + ig [A_\a, B^m]),}
$$F_{mn} = \p_{[m} B_{n]} + ig [B_m, B_n] ={1\over 8} (\g_{mn})_\a{}^\b
(D_\b W^\a + ig \{A_\b, W^\a\}). $$
Under the gauge transformation of \superg, 
\eqn\gauget{\d B_m = \p_m \Lambda + ig [B_m,\Lambda],\quad
\d W^\a = ig [W^\a,\Lambda],\quad \d F^{mn} = ig [F^{mn},\Lambda].}

To show that $A_\a(x,\t)$ describes on-shell super-Yang-Mills theory,
it will be useful to first note that in ten dimensions
any symmetric bispinor $f_{\a\b}$ can be decomposed in terms
of a vector and a five-form as
$f_{\a\b} = \g^m_{\a\b} f_m + \g^{mnpqr}_{\a\b} f_{mnpqr}$
and any antisymmetric bispinor $f_{\a\b}$ can be decomposed in terms
of a three-form as $f_{\a\b} = \g^{mnp}_{\a\b} f_{mnp}.$
Since $\{D_\a,D_\b\}= i\g^m_{\a\b}\p_m$, one can check that 
$\d A_\a =  D_\a \Lambda + ig[A_\a,\Lambda]$ 
is indeed a gauge invariance of \eom.

Using $\Lambda(x,\t) = h_\a(x) \t^\a + j_{\a\b} (x)\t^\a \t^\b,$ one can
gauge away $(A_\a(x))|_{\t=0}$ and the
three-form part of $(D_\a A_\b(x))|_{\t=0}$. Furthermore, equation
\eom\ implies that the five-form part of 
$(D_\a A_\b(x))|_{\t=0}$ vanishes. So the lowest non-vanishing
component of $A_\a(x,\t)$ in this gauge is the vector component
$(D\g_m A(x))|_{\t=0}$ which will be defined as
$8i a_m(x)$. Continuing this type of argument to
higher order in $\t^\a$, one finds that there exists a gauge choice
such that
\eqn\compa{A_\a(x,\t) ={i\over 2}
(\g^m \t)_\a a_m(x) + {i\over{12}}(\t\g^{mnp}\t) (\g_{mnp})_{\a\b} 
\chi^\b(x) +  ... }
where $a_m(x)$ and $\chi^\b(x)$ are $SO(9,1)$ vector and spinor fields 
satisfying \lorentz\
and where the component fields in $...$ are functions of spacetime
derivatives of $a_m(x)$
and $\chi^\b(x)$.
Furthermore, this gauge choice leaves the residual gauge transformations
of \gaui\ where $s(x) = (\Lambda(x))|_{\t=0}$. Also,
one can check that the $\t=0$ components of the superfields
$B_m$, $W^\a$ and $F_{mn}$ of \fs\
are $a_m$, $\chi^\a$ and $f_{mn}$ respectively.
So the equations of motion and gauge invariances of \eom\ and
\superg\ correctly
describe on-shell super-Yang-Mills theory.

One would now like to obtain this super-Poincar\'e covariant description
of super-Yang-Mills
theory by quantizing the superparticle. As will now be shown,
this can be done by constructing a BRST-like operator out of the
fermionic constraints $d_\a$ of \dirac.

\newsec{ Covariant Quantization of the Superparticle}

In the case of Chern-Simons theory, the gauge transformation
$\d A_\mu = \p_\mu\Lambda$ was generated by the constraints $P_\mu$.
So for the superparticle, the gauge transformation $\d A_\a = D_\a\Lambda$
suggests using the constraints $d_\a$.
However, the constraints $d_\a$ are not
all first-class, so 
\eqn\brst{Q = \l^\alpha d_\alpha}
would not
be a nilpotent operator for generic $\l^\a$. 
However, since \antic\ implies that 
$Q^2= (\l^\a d_\a)^2 = \half\l^\alpha \l^\beta 
\gamma^m_{\alpha \beta} P_m$, $Q$ would be nilpotent if
$\l^\alpha$ satisfied the condition
\eqn\pure{ \l^\a \g^m_{\a\b} \l^\b =0}
for $m=0$ to 9. The condition of \pure\ is the definition
of a pure spinor \ref\howe{
P. Howe, Phys. Lett. B258 (1991) 141.}
and, as will now be shown,
implies that only eleven components of
$\l^\a$ are independent parameters.

\subsec{Pure spinors}

To solve the constraint of \pure, it is convenient to first Wick-rotate
to Euclidean space and write the $SO(10)$ spinor $\l^\a$ using
$SU(5)$ notation as $\l^{\pm\pm\pm\pm\pm}$ where $\pm$
denotes if the component is annihilated by
$(\g_{2a-2}+i\g_{2a-1})$ or 
$(\g_{2a-2}-i\g_{2a-1})$ for $a=1$ to 5. For example, the component
$\l^{+-+-+}$ is annihilated by
$\g_0+i\g_1$,
$\g_2-i\g_3$,
$\g_4+i\g_5$,
$\g_6-i\g_7$,
and
$\g_8+i\g_9$. For a sixteen-component Weyl spinor, $\l^{\pm\pm\pm\pm\pm}$
contains either five $+$'s, three $+$'s, or one $+$, which transform
respectively under $SU(5)$ as $1$, $\overline{10}$, and $5$
representations. These $SU(5)$ representations will be called
$\l^+$, $\l_{ab}$ and $\l^a$ where $a=1$ to 5 and $\l_{ab}=-\l_{ba}$.
One can check that any pure spinor $\l^\a$ satisfying \pure\ 
can be parameterized as
\eqn\param{\l^+ = \g,\quad \l_{ab} = \g u_{ab},\quad \l^a = -{\g\over 8}
\e^{abcde} u_{bc} u_{de}}
where $u_{ab}= -u_{ba}$ and $\g$ are eleven independent parameters. To show
that \param\ satisfies \pure, note that
\eqn\notet{\l\g^a\l = \l^+\l^a +{1\over 8}\e^{abcde}\l_{bc}\l_{de}, \quad
\l\g_a\l = \l^b \l_{ab}}
where $\g^a = {1\over{\sqrt{2}}}
(\g_{2a-2}-i\g_{2a-1})$ and
$\g_a = {1\over{\sqrt{2}}}
(\g_{2a-2}+i\g_{2a-1})$. Also note that the parameterization of \param\
is singular when $\l^+=0$ since 
$\g\to 0$ and $u_{ab}\to\infty$ when $\l^+\to 0$.

To obtain the BRST operator of \brst\ from gauge-fixing, start with
the worldline action
\eqn\su{S=\int d\tau (\dot x^m  P_m -\half P^m P_m
+ \dot \theta^\alpha p_\a + 
\half\dot u_{ab} v^{ab} + l  (\l^\a/\g) d_\a)}
where $p_\a$ is the conjugate momentum for $\t^\a$, $v^{ab}$
is the conjugate momentum for $u_{ab}$, $\l^\a/\g$ is defined in
terms of $u_{ab}$ using the definition of \param, 
and $l$ is the Lagrange multiplier for the constraint $(\l^\a/\g) d_\a$.
\foot{Note that the term $-\half P^m P_m$ appears before gauge-fixing,
implying that the action of \su\ is not invariant under
worldline reparameterizations. This fact is probably related to the indirect
manner in which BRST invariance imposes the mass-shell condition.}
After gauge fixing $l=0$, one obtains the action
\eqn\st{S=\int d\tau (\dot x^m  P_m-\half P^m P_m
+ \dot \theta^\alpha p_\a + 
\half \dot u_{ab} v^{ab} + \dot\g \b)}
with the BRST operator $Q=\l^\a d_\a$ where
$(\g,\b)$ are the bosonic Fadeev-Popov ghost and antighost for the
constraint $(\l^\a/\g) d_\a$.

Note that
the action and BRST operator of \st\ and \brst\ are
spacetime supersymmetric since
$\dot x^m P_m +\dot\t^\a p_\a =\Pi^m P_m +\dot\t^\a d_\a$
where $\Pi^m$ and $d_\a$ are defined in \defpi\ and \dirac.
Furthermore, although $(\g,\b)$ and $(u_{ab},v^{ab})$ do not
transform linearly under Lorentz transformations, one can define
Lorentz generators such that the pure spinor $\l^\a$ of
\param\ does transform linearly as $\d \l^\a = \half (\g^{mn})^\a{}_\b \l^\b$
under the transformation generated by $N^{mn}$.
These Lorentz generators are given by
\eqn\lorentzg{N^{ab} = v^{ab},\quad N_{ab} = u_{ac} u_{bd} v^{cd} 
- u_{ab}\g\b,}
$$N^a_b = u_{bc} v^{ac} - {1\over 5} \d^a_b u_{cd} v^{cd},\quad
N = {5\over 2} \g\b -u_{ab} v^{ab},$$
where the 45 $SO(10)$ Lorentz generators $N^{mn}$ have been decomposed in
terms of their irreducible $SU(5)$ representations 
$(N^{ab},N_{ab},N^a_b,N)$ which transform as
$(10,\overline{10},24,1)$ representations.
So the action and BRST operator of \st\ and \brst\ are super-Poincar\'e
invariant and it will now be shown that they correctly describe
super-Yang-Mills theory.

\subsec{Quantization}

The most general super-Poincar\'e covariant wavefunction that can
be constructed from
$(x^m,\t^\a,\l^\a)$ is\foot{It will be assumed that $\Psi$ transforms
covariantly under Lorentz transformations, which implies that it only
depends on $\g$ and $u_{ab}$ through the combination $\l^\a$ of \param.}
\eqn\spf{\Psi(x,\t,\l) = C(x,\t) + \l^\a A_\a(x,\t) + 
(\l\g^{mnpqr}\l) A^*_{mnpqr}(x,\t) + \l^\a \l^\b \l^\g C^*_{\a\b\g}(x,\t) 
+ ...}
where $...$ includes superfields with more than three powers of $\l^\a$.
Since $Q\Psi = -i\l^\a D_\a C -i \l^\a \l^\b D_\a A_\b + ...$,
$Q\Psi=0$ implies that $A_\a(x,\t)$ satisfies the equation of motion
$\l^\a \l^\b D_\a A_\b=0$. But since 
$\l^\a \l^\b$ is proportional to $(\l\g^{mnpqr}\l) \g_{mnpqr}^{\a\b}$, 
this implies
that $D\g^{mnpqr}A=0$, which is the linearized version of the
super-Yang-Mills
equation of motion of \eom. Furthermore, if one defines the
gauge parameter $\Omega= i\Lambda + \l^\a \omega_\a + ...$, the
gauge transformation $\d\Psi = Q\Omega$ implies $\d A_\a = D_\a \Lambda$
which is the linearized super-Yang-Mills gauge transformation  
of \superg.

As was shown in \ref\cohom{N. Berkovits, JHEP 09 (2000) 046,
hep-th/0006003.}, the only states at non-zero momentum in the
cohomology of $Q$ are the on-shell super-Yang-Mills gluon and
gluino, $a_m(x)$ and $\chi^\a(x)$, and their antifields,
$a^{*m}(x)$ and $\chi^*_\a(x)$ \foot{The presence of the antifields
can be seen from the doubling of the 
cohomology at ghost-numbers $+1$ and $+2$.}.
Since gauge invariances of the
antifields correspond to equations of motion of the fields and
vice versa, one expects $a^{*m}$ and $\chi^*_\a$ to satisfy
the linearized equations of motion $\p_m a^{*m}=0$ with the linearized
gauge invariances
\eqn\afeom{\d a^{*m} = \p_n (\p^n s^m - \p^m s^n),\quad
\d\chi_\a^* = \g^m_{\a\b}\p_m \kappa^\b}
where $s^m $ and $\kappa^\b$ are gauge parameters.

The fields $a_m$ and $\chi^\a$ appear in components of
$A_\a$ as in \compa, and
the antifields $a^{*m}$ and $\chi^*_\a$ appear
in components of the
ghost-number $+2$ superfield $A^*_{mnpqr}$ of \spf.
Using $Q\Psi=0$ and $\d\Psi = Q\Omega$, $A^*_{mnpqr}$
satisfies the linearized
equation of motion $\l^\a(\l\g^{mnpqr}\l) D_\a A^*_{mnpqr}=0$
with the linearized
gauge invariance $\d A^*_{mnpqr} = \g_{mnpqr}^{\a\b} D_\a \omega_\b$.
Expanding $\omega_\a$ and $A^*_{mnpqr}$ in components, one learns
that 
$A^*_{mnpqr}$ can be gauged to the form
\eqn\astar{A^*_{mnpqr} = (\t\g_{[mnp}\t)(\t\g_{qr]})^\a \chi^*_\a(x) +
(\t\g_{[mnp}\t)(\t\g_{qr]s}\t) a^{*s}(x) + ...}
where $\chi^*_\a$ and $a^{*s}$ satisfy the equations of motion 
and residual gauge invariances of \afeom, and $...$ involves
terms higher order in $\t^\a$ which depend on derivatives of
$\chi^*_\a$ and $a^{*s}$.

As will be shown in the appendix, there are also zero momentum
states in the cohomology of $Q$. In addition to the states described
by the zero-momentum gluon, gluino, antigluon, and antigluino, there
are also zero-momentum ghost and antighost states $c$ and $c^*$
in the $\t=0$ component of the ghost-number zero
superfield, $C(x,\t) = c(x) + ...$, and in the $(\t)^5$ component of the
ghost-number $+3$ superfield,
$C^*_{\a\b\g}(x,\t)= ... + c^*(x) (\g^m\t)_\a (\g^n\t)_\b
(\g^p\t)_\g (\t\g_{mnp}\t) + ... .$ 
So although $\Psi$ of \spf\ contains superfields of arbitrarily high
ghost number, only superfields with ghost-number between zero and three
contain states in the cohomology of $Q$.

The linearized equations of motion and gauge invariances $Q\Psi=0$
and $\d\Psi =Q\Omega$ are easily generalized to the non-linear
equations of motion and gauge invariances
\eqn\nonls{Q\Psi + g\Psi \Psi =0,\quad \d\Psi = Q\Omega + g[\Psi,\Omega]}
where $\Psi$ and $\Omega$ transform in the adjoint representation of
the gauge group. For the superfield $A_\a(x,\t)$, \nonls\ implies
the super-Yang-Mills equations of motion and gauge transformations of
\eom\ and \superg.
Furthermore, the equations of motion and gauge transformation of \nonls\
can be obtained from the spacetime action\foot
{This spacetime action was first proposed to me by John Schwarz
and Edward Witten. Because the action only involves integration over
five $\t$'s, it is not manifestly spacetime supersymmetric. Nevertheless,
the equations of motion coming from this action have the same physical
content as the manifestly spacetime supersymmetric equations of motion
$Q\Psi + g\Psi\Psi=0$. This is because all components in
$Q\Psi + g\Psi\Psi=0$ with more than five $\t$'s are auxiliary equations
of motion. So removing these equations of motion only changes auxiliary
fields to gauge fields but does not affect the physical content of the
theory. Unfortunately, this does not seem to be true after including
the massive modes of the superstring. So there does not appear to
exist a cubic superstring field theory action which reproduces the 
equations of motion
$Q\Psi + g\Psi\times\Psi=0$ where $\Psi$ is the superstring field and
$\times$ is Witten's midpoint interaction.}
\eqn\yma{{\cal S}= Tr\int d^{10}x \langle \half\Psi Q \Psi + 
{g\over 3}\Psi\Psi\Psi\rangle}
using the normalization definition that 
\eqn\norm{\langle (\l\g^m\t)(\l\g^n\t)(\l\g^p\t)(\t\g_{mnp}\t) \rangle =1.}
Although \norm\ may seem strange, it resembles the normalization of
\comp\ in that $\langle \Psi\rangle = c^*(x)$ where $c^*(x)$ is the
spacetime antighost. 
\foot{It would be interesting to try to derive \nonls\
from an eleven-dimensional action, in analogy to the four-dimensional
Chern-Simons action of footnote 3. One natural guess would be to extend
$\lambda$ and $\theta$ to eleven-dimensional spinors and define
the non-vanishing normalization as $\langle
(\l\g^m\t)(\l\g^n\t)(\l\g^p\t)(\l\g^q\t)(\t\g_{mnpq}\t)\rangle =1$.} 
After writing \yma\ in terms of component fields
and integrating out auxiliary fields, it should be possible to show
that \yma\ reduces to the standard Batalin-Vilovisky action for
super-Yang-Mills,
\eqn\bvym{{\cal S} = Tr\int d^{10}x ({1\over 4} f_{mn} f^{mn}
+\chi^\a \g^m_{\a\b} (\p_m\chi^\b +ig[a_m,\chi^\b])}
$$+ i a^{*m}(\p_m c + ig[a_m,c]) -g \chi^*_\a \{\chi^\a,c\} -g c c c^*).$$

\subsec{Non-abelian super-Yang-Mills background}

As in Chern-Simons theory, one can modify the worldline action 
of \st\ to describe the superparticle in a non-abelian super-Yang-Mills
background
with $SO(N)$ (or $U(N)$) gauge group by including $N$ real (or complex)
fermions. For $SO(N)$ gauge group, the worldline action is
\eqn\was{S=\int d\tau (\dot x^m P_m -\half P^m P_m
+ \dot\t^\a p_\a -{i\over 2}
\eta_I\nabla\eta_I + \half
\dot u_{ab} v^{ab} + \dot\g \b)}
where 
\eqn\vert{\nabla\eta_I = \dot \eta_I + g\eta_J
(\dot\t^\a \tilde A_\a^{IJ} + \Pi^m \tilde B_m^{IJ} +
d_\a \tilde W^{\a~IJ} +\half N^{mn} \tilde F_{mn}^{IJ}),}
$\tilde A_\a^{IJ}$ is the background super-Yang-Mills gauge field,
$\tilde B_m^{IJ}$,
$\tilde W^{\a~IJ}$ and $\tilde F_{mn}^{IJ}$ are background superfields
constructed from $\tilde A_\a^{IJ}$ as in \fs,
$\Pi^m$ is defined in \defpi, and $N_{mn}$ is defined in \lorentzg.

As will now be shown, the coupling of $\tilde A_\a^{IJ}$ has been
chosen such that $\l^\a d_\a$ is conserved. To show this, note that
$\dot \l^\a =-{{ig}\over 8} \eta_I\eta_J (\g^{mn}\l)^\a F_{mn}^{IJ}$ and
\eqn\dotda{\dot d_\a = {{ig}\over 2}{{\p}\over{\p\tau}}(\eta_I\eta_J
\tilde A_\a^{IJ}) -{{ig}\over 2}\eta_I\eta_J
(-\dot\t^\b D_\a \tilde A^{IJ}_\b
+\Pi^m D_\a \tilde B_m^{IJ} }
$$- d_\b D_\a\tilde W^{\b~IJ} + \half N_{mn} D_\a\tilde F_{mn}^{IJ}
+ \g^m_{\a\b} (i\dot\t^\b \tilde B_m^{IJ} +\Pi_m \tilde W^{\b~IJ})$$
$$= {{ig}\over 2}\eta_I\eta_J (\dot\t^\b (D_{(\a} \tilde A_{\b)}^{IJ} +g
\tilde A_{(\a}^{IK} \tilde A_{\b)}^{KJ} -i\g^m_{\a\b} \tilde B_m^{IJ})
+\Pi^m (-(\nabla_\a \tilde B_m)^{IJ} + \p_m \tilde A_\a^{IJ} + 
\g_{m~\a\b} W^{\b~IJ})$$
$$+ d_\b (\nabla_\a \tilde W^\b)^{IJ}
-\half N^{mn} (\nabla_\a\tilde F_{mn})^{IJ}),$$
where $(\nabla_\a s)^{IJ} = D_\a s^{IJ} + g (s^{IK} \tilde A_\a^{KJ} -
s^{JK} \tilde A_\a^{KI})$ and the equations of motion
$\dot\t^\a=-{{ig}\over 2}\eta_I\eta_J \tilde W^{\a~IJ}$ and
$P^m=\Pi^m$ have been used.
So using the definitions of \fs, 
\eqn\cons{{{\p}\over {\p\tau}}(\l^\a d_\a) = -{{ig}\over 4}\eta_I\eta_J
\l^\a N^{mn} 
(\nabla_\a \tilde F_{mn})^{IJ}.} 
But $F_{mn}^{IJ}$ satisfies 
$ (\nabla_\a \tilde F_{mn})^{IJ} ={i\over 8}
\g_{\a\b ~[m}(\nabla_{n]} W^\b)^{IJ}$ where
$(\nabla_n s)^{IJ} = D_n s^{IJ} + g (s^{IK} \tilde B_n^{KJ} -
s^{JK} \tilde B_n^{KI})$.
Also, one can check that $N^{mn} =\half \l\g^{mn} w$ where
$w_\a$ is an anti-Weyl spinor with $SU(5)$ components
\eqn\defw{w_+=\b,\quad w^{ab} = v^{ab},\quad w_a=0.}
So 
\eqn\ver{\l^\a N^{mn} 
(\nabla_\a \tilde F_{mn})^{IJ} 
={i\over{16}}
\l^\a (\l\g^{mn}w) \g_{\a\b~[m}(\nabla_{n]} \tilde W^\b)^{IJ}
= -{i\over 8}
(\l^\delta w_\delta) \l^\a\g^n_{\a\b} (\nabla_n \tilde W^\b)^{IJ} }
using \pure\
and the gamma-matrix identity that $\eta_{mn}\g^m_{(\a\b}\g^n_{\d)\k}=0$.
But one now can use that $\g^n_{\a\b}(\nabla_n\tilde W^\b)^{IJ}=0$
to imply that \ver\ vanishes, and therefore $\l^\a d_\a$ is conserved.

Finally, it will be shown that $Q(\eta_I\eta_J\Psi^{IJ})=0$
gives the correct equation of motion in the presence of
the background gauge field $\tilde A_\a^{IJ}$.
Since the canonical momentum
for $\t^\a$ and $x^m$ are
\eqn\canm{{{\p L}\over{\p\dot\t^\a}}=p_\a -{{ig}\over 2}\eta_I\eta_J(
\tilde A_\a^{IJ} -{i\over 2}\t^\b\g^m_{\a\b} \tilde B_m^{IJ}),}
$${{\p L}\over{\p\dot x^m}}= P_m -{{ig}\over 2}\eta_I\eta_J
\tilde B_m^{IJ},$$
one finds that 
\eqn\covb{Q = \l^\a d_\a = 
\l^\a (p_\a +{i\over 2}P_m \g^m_{\a\b}\t^\b )}
$$= \l^\a
({{\p L}\over{\p\dot\t^\a}} +{i\over 2}
{{\p L}\over{\p\dot x^m}}\g^m_{\a\b}\t^\b 
+{{ig}\over 2}\eta_I\eta_J\tilde A_\a^{IJ}).$$
So 
\eqn\covq{Q(\eta_I\eta_J\Psi^{IJ})= -i\eta_I\eta_J (\l^\a (\nabla_\a C)^{IJ}
+ \l^\a \l^\b (\nabla_\a A_\b)^{IJ} + \l^\a(\l\g^{mnpqr}\l)(\nabla_\a
A^*_{mnpqr})^{IJ} + ...)}
where $(\nabla_\a s)^{IJ} = D_\a s^{IJ} + g (s^{IK} \tilde A_\a^{KJ} -
s^{JK} \tilde A_\a^{KI})$,
which correctly covariantizes the supersymmetric derivatives with respect
to the background gauge field. 

\newsec{Appendix: Superparticle Cohomology at Zero Momentum}

In this appendix, the zero momentum cohomology of $Q=\l^\a d_\a$
will be computed for arbitrary ghost number and shown
to correspond to the ghost, gluon, gluino, antigluino, antigluon,
and antighost of super-Yang-Mills. Since $Q=\l^\a p_\a$
when $P_m=0$, the only reason one has non-trivial cohomology is because
$\l^\a$ is constrained by \pure. It will now be proven that 
the cohomology of $Q=\l^\a p_\a$ with constrained $\l^\a$ is
equivalent to the ``linear'' cohomology of $\widehat Q$ with
unconstrained $\l^\a$ where
\eqn\defwq{\widehat Q = \l^\a p_\a + (\l\g^m\l) b_m 
+ c^m (\l\g_m f) + (\l\g_m\l)(j\g^m g) - 2(j_\a \l^\a)(g_\b\l^\b)}
$$
+ (k\g_m\l) r^m + (\l\g^m\l) s_m t, $$
and ``linear'' cohomology signifies elements in the cohomology of
$\widehat Q$ which are at most linearly dependent on the new variables
$(c^m, g_\a,k^\a,s_m,u)$. 
Note that
$(c^m,b_m)$, $(g_\a,f^\a)$, $(k^\a,j_\a)$, $(s_m,r^m)$, 
and $(u,t)$
are pairs of new variables and their conjugate momentum which
have been added to the Hilbert space.
The pairs $(c^m,b_m)$, $(k^\a,j_\a)$ and $(u,t)$ are fermions of
ghost-number $(1,-1)$, $(2,-2)$ and $(3,-3)$ respectively,
and the pairs $(g_\a,f^\a)$ and $(s_m,r^m)$ are bosons of ghost-number
$(1,-1)$ and $(2,-2)$ respectively.

\subsec{Equivalence of $Q$ and $\widehat Q$ cohomologies}

To relate the cohomologies of $Q$ and $\widehat Q$, consider a
state $F(\l,\t)$ in the cohomology of $Q$ with constrained $\l^\a$.
Then $QF= (\l\g^m\l)\tau_m$ for some $\tau_m$. So 
$\widehat Q(F- c^m\tau_m) = c^m Q\tau_m$. But $Q^2 F=0$ implies that
$(\l\g^m\l) Q\tau_m=0$, which implies that 
$Q\tau_m = \l\g_m\psi$ for some $\psi^\a$. So
$\widehat Q(F- c^m\tau_m -g_\a\psi^\a) = -g_\a Q\psi^\a$. 
But $Q^2 \tau_m=0$ implies that $\l\g_m Q\psi=0$, which
implies that $Q\psi^\a = (\l\g^n\l)(\g_n\rho)^\a -2 \l^\a (\l^\b \rho_\b)$
for some $\rho_\b$. This line of argument continues until one has
$\widehat Q(F- c^m\tau_m -g_\a\psi^\a + k^\a \rho_\a + s_m \s^m + u\k) 
= u Q\kappa$. Finally, $Q^2 \s^m
=0$ implies that $(\l\g^m\l)Q\kappa =0,$ 
which implies that $Q\kappa=0$. So for any 
state $F(\l,\t)$ in the cohomology of $Q$ with constrained $\l^\a$, one
can construct a state $\widehat F$ annihilated by $\widehat Q$ 
which is at most linear in
$(c^m, g_\a,k^\a,s_m,u)$. 

To show that $\widehat F$ is in the cohomology of $\widehat Q$,
suppose that $\widehat F=\widehat Q \widehat \Omega$ for
some $\widehat \Omega = \Omega(\l,\t) + c^m \xi_m(\l,\t) + ...$.
Then since $F$ is the term in $\widehat F$ which is independent
of the new variables, $F= Q\Omega + (\l\g^m\l)\xi_m$. But
this is not possible if $F$ is in the cohomology of $Q$ with
constrained $\l^\a$, so 
$\widehat F\neq\widehat Q \widehat \Omega$ for any $\widehat\Omega$.

Now suppose that one starts with a state $\widehat F$ in the cohomology
of $\widehat Q$ which is at most linear in
$(c^m, g_\a,k^\a,s_m,u)$, i.e.
\eqn\deff{\widehat F = F(\l,\t) + c^m \tau_m(\l,\t) + g_\a\psi^\a(\l,\t)
+ k^\a \rho_\a(\l,\t) +s_m\s^m(\l,\t) + u \kappa(\l,\t).}
Then $\widehat Q\widehat F = 0$ implies that 
$QF = -(\l\g^m\l) \tau_m$, so $F$ is annihilated by $Q$ with constrained
$\l^\a$.

To show that $F$ is in the cohomology of $Q$, suppose that 
$F= Q\Omega + (\l\g^m\l)\xi_m$ for some $\Omega(\l,\t)$ and $\xi_m(\l,\t)$.
Then $\widehat F 
= \widehat Q (\Omega + c^m \xi_m) + c^m (\tau_m + Q\xi_m) + g_\a\psi^\a + ...$. 
So $\widehat Q\widehat F =0$ implies that
$(\l\g^m\l)(\tau_m + Q\xi_m)=0$, which implies that 
$\tau_m + Q\xi_m=\l\g^m\chi$ for some $\chi^\a$. So
$\widehat F = \widehat Q(\Omega+ c^m\xi_m + g_\a\chi^\a) + 
g_\a(\psi^\a -Q\chi^\a) + k^\a \rho_\a + ...$. This argument continues
until one finds that 
$\widehat F = \widehat Q (\Omega + c^m\xi_m + g_\a\chi^\a 
+ ... + u\varepsilon)$, which is not possible if $\widehat F$ is in
the cohomology of $\widehat Q$. So it has been proven that the
cohomology of $Q$ with constrained $\l^\a$ is equivalent to the ``linear''
cohomology of $\widehat Q$ with unconstrained $\l^\a$. 

\subsec{Evaluation of $\widehat Q$ cohomology}

Since $\widehat Q = \l^\a \widehat p_\a$ where $\l^\a$ is unconstrained and
$$\widehat p_\a = p_\a + (\g^m\l)_\a b_m + 
c^m (\g_m f)_\a + (\g_m\l)_\a(j\g^m g) - 2j_\a (g_\b\l^\b)
+ (\g_m k)_\a r^m + (\g^m\l)_\a s_m t ,$$
it is easy to evaluate the cohomology of $\widehat Q$. Using the
quartet mechanism, one can choose a gauge such that states in
the cohomology are independent
of $\l^\a$ and $\t^\a$. So states in the ``linear'' cohomology are
represented by the elements $(1,c^m,g_\a,k^\a,s_m,u)$, which have
ghost-number $(0,1,1,2,2,3)$ respectively. 

To relate these elements to states in the cohomology of $Q$,
one needs to find gauge-invariant version of these elements
which commute with $\widehat Q$. For example, $c^m - i\l\g^m\t$
and $g_\a -i (\t\g_m)_\a c^m +{2\over 3}(\l\g_m\t)(\g^m\t)_\a$ 
commute with $\widehat Q$, so $\l\g^m\t$ and $(\l\g_m\t)(\g^m\t)_\a$
are the states in the cohomology of $Q$ which are associated with
$c^m$ and $g_\a$. Similarly, one can show that 
$(\l\g^m\t)(\l\g^n\t)(\t\g_{mn})^\a$ is the state associated with $k^\a$,
$(\l\g^m\t)(\l\g^n\t)(\t\g_{mnp}\t)$ is the state associated with $s_p$,
and $(\l\g^m\t)(\l\g^n\t)(\l\g^p\t)(\t\g_{mnp}\t)$ is the state
associated with $u$.

Comparing these states with
the superfields in \spf, one finds
that the zero momentum states in the cohomology
of $Q$ correspond to the ghost, gluon, gluino, antigluino, antigluon, and
antighost of super-Yang-Mills.

\vskip 15pt

{\bf Acknowledgements:}
I would like to thank Warren Siegel for
useful discussions, Jo\~ao Barcelos-Neto for suggesting that 
these notes be written up, and
the Clay Mathematics Institute, 
CNPq grant 300256/94-9, Pronex 66.2002/1998-9, 
and FAPESP grant 99/12763-0 for partial financial support.
This research was partially conducted during the period the author
was employed by the Clay Mathematics Institute as a CMI Prize Fellow.

\listrefs

\end